\documentclass[letterpaper,12pt]{article}
\pdfoutput=1
\usepackage{jheppub}
\usepackage{epsfig}
\usepackage{amsmath}
\usepackage{subfigure}

\DeclareSymbolFont{AMSa}{U}{msa}{m}{n}
\DeclareSymbolFont{AMSb}{U}{msb}{m}{n}
\let\Box\relax
\DeclareMathSymbol{\Box}{\mathord}{AMSa}{"03}

\newcommand{\be}{\begin{equation}}
\newcommand{\ee}{\end{equation}}
\newcommand{\bea}{\begin{eqnarray}}
\newcommand{\eea}{\end{eqnarray}}

\newcommand{\f}{\frac}

\newcommand{\ds}{\displaystyle}

\newcommand{\Od}{{\mathcal{O}}}
\newcommand{\Rd}{{\mathcal{R}}}

\title{Non-Gaussianity from Excited Initial Inflationary States}

\author{Aditya Aravind,}
\author{Dustin Lorshbough,}
\author{and Sonia Paban}
\affiliation{Department of Physics and Texas Cosmology Center\\ The University of Texas at Austin,
TX 78712.}
\emailAdd{Aditya@physics.utexas.edu}
\emailAdd{Lorsh@utexas.edu}
\emailAdd{Paban@zippy.ph.utexas.edu}

\abstract{We study squeezed limit $\ds{f_{NL}}$ generation by excited initial inflationary states in a model independent way.  We restrict ``excited" to mean a Bogoliubov transformation of the Bunch Davies state.  We simultaneously impose the constraints that the observable power spectrum is nearly scale invariant over at least three decades and that the observable modes today be subhorizon at the beginning of inflation while not causing significant backreaction.  We show that most excited initial inflationary states for single field inflationary models with negligible superhorizon evolution do not produce an observable squeezed limit $\ds{f_{NL}}$.  The case in which one mode is in the Bunch Davies state while the other two modes are in an excited state with $\ds{0.01<|\beta_k|\leq0.1}$ may generate a squeezed limit $\ds{f_{NL}}$ which is detectable with future experiments.}


\begin{document}
\maketitle
\flushbottom
\section{Introduction}
The standard single field inflationary scenario whereby a single scalar field undergoes slow roll inflation has been remarkably consistent with observational data\cite{WMAP}.  The usual assumption is that the fluctuations that we observe in our universe today were initially in a Bunch Davies state.  However, recently there have been attempts to study models of inflation which have an initial state which is related to the standard Bunch Davies vacuum through a Bogoliubov transformation\cite{Greene05,Chen07,Holman08,Agullo11,Ganc11,Agarwal12,Carney12,Chialva12,Kundu12,Gong13}.  We are restricting our usage of the word ``excited" to mean only states which have a well-defined non-zero Bogoliubov coefficient $\ds{\beta_k}$.  We also are not considering models with superhorizon evolution such as those studied in \cite{Namjoo11,Chen12}. It was shown in \cite{Dey12} that pre-inflation background anisotropy models, in the so called non-planar limit, result in an excited initial state with rescaled Bogoliubov coefficients given 
in equation (3.9) of that paper.  Therefore these models should be constrained by our analysis as well, the only possible difference being that the Bogoliubov coefficients depend on a special direction.  We will show most models of inflation with excited initial states and negligible superhorizon evolution cannot consistently result in an observably large squeezed limit $\ds{f_{NL}}$.\\

In section 2 we prove that in general, based upon the observation of three decades of nearly scale invariant power spectrum, most excited initial states cannot be observationally distinguished from the standard Bunch Davies state by the measurement of squeezed limit $\ds{f_{NL}}$.  We will consider both the mixed case whereby today's observable modes are composed of both Bunch Davies modes and excited modes in addition to the unmixed case whereby today's observable modes are composed solely of the latter.  In order to make our argument less abstract, we explicitly demonstrate this for the interesting case of large excitations in section 3.  We will begin section 3 by reviewing an excited state scenario in the literature that is physically motivated by the introduction of an initially kinetic dominated era prior to the inflationary era \cite{Carney12}.  We then explicitly show that in this model the squeezed limit $\ds{f_{NL}}$ for the case in which all modes under consideration are in the excited initial 
state is not observable. In section 4 we conclude.
\section{Excited Initial State $\ds{f_{NL}}$ Undetectability}
In order to study the squeezed limit $\ds{f_{NL}}$ for single field inflationary scenarios in a model independent fashion, we will simultaneously impose four constraints:\footnote{Also see Raphael Flauger's presentation at the workshop “Critical Tests of Inflation Using Non-Gaussianity” at http://www.mpa-garching.mpg.de/$\sim$komatsu/meetings/ng2012/.}\\
(1) Modes that we observe today were subhorizon at the beginning of inflation.\\
(2) Slow roll inflation occurred undisturbed by fluctuation backreaction.\\
(3) Any superhorizon evolution was negligible.\\
(4) There are at least three decades of nearly scale invariant observable power spectrum.\\

In sections 2.1 and 2.2 we impose these constraints to show that the only scenario which may result in an order unity $f_{NL}$ is the case in which one mode is in the Bunch Davies state while the two modes are in an excited state with $\ds{0.01<|\beta_k|\leq0.1}$.  In section 2.3 we will parameterize these constraints explicitly in order to demonstrate their robustness.
\subsection{Mixed Case: Bunch Davies Modes and Excited Modes}
For general excited states, the power spectrum of comoving curvature perturbations ($\Rd$) is given by
\begin{equation}
P_\Rd(k)=\f{1}{4k^3\epsilon}\f{H^2}{M_P^2}|\alpha_k+\beta_k|^2
\end{equation}
with $\ds{|\alpha_k|^2-|\beta_k|^2=1}$ (see for example, \cite{Ganc11}).  The near scale invariance of the observed power spectrum therefore prohibits the case of mixing Bunch Davies modes with excited modes unless $\ds{|\beta_k|\leq 0.1}$\cite{Greene05}.  The leading contribution to $\ds{f_{NL}}$ in the mixing case for small $\ds{|\beta_k|}$ is of the form $\ds{\epsilon|\beta_k|(k_1/k_3)}$ \cite{Ganc11} with $(k_1/k_3)\leq10^{3}$, where $k_1$ is the short wavelength mode and $k_3$ is the long wavelength mode.  This assumes that the two short wavelength modes are in an excited state and the third long wavelength mode is in the Bunch Davies state, otherwise one does not obtain an enhancement in the squeezed limit.  Constraining $\epsilon$ in terms of the tensor to scalar ratio, $r\leq0.13$\cite{WMAP}, we obtain\footnote{Here we have assumed that the tensor spectrum is left unchanged by setting the initial state of scalar perturbations to be something other than Bunch Davies.  Using an initially excited state 
for the tensor perturbations would further suppress $\epsilon$ and therefore further suppress $f_{NL}$ as we 
shall see.}
\begin{equation}\label{eq:epsilon}
\epsilon=\f{r}{16}|\alpha_k+\beta_k|^2.
\end{equation}
The constraint becomes $\epsilon\leq 10^{-2}$ for the mixing case of $|\beta_k|\leq0.1$, resulting in a maximum squeezed limit $\ds{f_{NL}}$ contribution, up to numerical coefficients, of order unity.  For $|\beta_k|\leq0.01$, however, we obtain $\ds{f_{NL}\sim10^{-1}}$ which is undetectably small.  Therefore the mixed case does not result in an observationally distinguishable squeezed limit $f_{NL}$ from the standard Bunch Davies result if $|\beta_k|<0.01$, but may for $0.01<|\beta_k|\leq0.1$.
\subsection{Unmixed Case: Only Excited Modes}
Now we restrict our attention to the case of all observable modes being initially in an excited state.  That the modes are observable today dictates that they were subhorizon at the beginning of inflation, $\ds{H< p}$, where $\ds{p}$ is the physical momentum.  Additionally, we require that the excited state modes do not generate enough backreaction to spoil slow roll inflation nor bring into question the validity of perturbation theory by imposing that the fluctuation energy density be much less than the total energy density, $\ds{\langle\rho_{\Rd}\rangle\ll H^2M_P^2}$.  Computing the energy density of the fluctuations explicitly using adiabatic subtraction and assuming that the Bogoliubov coefficients are approximately scale invariant\footnote{Except for in the case of $\ds{|\beta_k|\ll1}$ which is too small to have its 
scale dependence constrained by power spectrum observations, one may take any value for $\ds{|\beta|}$ 
that well represents the decades of interest and it will not alter our conclusion.}$^{,}$\footnote{This differs very slightly from the analysis conducted in \cite{Holman08,Ganc11} whereby it was assumed the coefficient would 
exponentially decay in momentum.  However this difference will not invalidate our discussion of backreaction since the modes whose contribution survives adiabatic subtraction have an essentially 
constant exponential term which is near unity.} ($\ds{|\beta_k|\sim|\beta|}$) gives a leading term which is quartic in momentum,
\begin{equation}
\begin{array}{lll}
\left\langle \rho_{\mathcal{R}} \right\rangle&=&\ds{\epsilon\int_0^{k_{\text{UV}}}\f{d^3k}{(2\pi)^3}\left[|\dot{R}_k|^2-|\dot{f}_{\text{BD,k}}|^2+\left(\f{k}{a}\right)^2\left(|R_k|^2-|f_{\text{BD,k}}|^2\right)\right]}\\
&=&\ds{\f{|\beta|^2}{8\pi^2}\left[p_{\text{
UV}}^4+\Od\left(H^2p_{\text{UV}}^2\right)\right].}\end{array}
\end{equation}
Here, $\ds{p_{UV}}$ is the highest physical momentum mode which is in an initially excited state and $\ds{R_k}$ is the comoving curvature perturbation mode function related to the Bunch Davies mode function $\ds{f_{\text{BD,k}}}$ through the standard Bogoliubov transformation.  The upper bound on the physical momentum is then given by $\ds{p\ll \sqrt{ 
10HM_P/|\beta|}}$.  Therefore combining our subhorizon lower bound with our backreaction upper bound we restrict the range of allowed momenta modes to be,
\begin{equation} \label{eq:AllowedMomenta}
H< p\ll \sqrt{10HM_P/|\beta|}.
\end{equation}

The range of observable modes today for which scale invariance holds spans at least three decades\cite{WMAP,Peiris10} and therefore we require the conservative case that there be at least three decades of allowed physical momenta range,
\begin{equation} \label{eq:ThreeDecades}
10^3\ll \f{\sqrt{10HM_P/|\beta|}}{H}.
\end{equation}

However the observable amplitude of the power spectrum \cite{WMAP} constrains the Hubble parameter H in terms of the slow roll parameter $\ds{\epsilon}$ and Bogoliubov coefficients $\ds{\alpha_k}$ and $\ds{\beta_k}$, $\ds{\left(M_P/H\right)^2<10^{7}\epsilon^{-1}|\alpha+\beta|^2}$.\footnote{$\ds{\Delta_\Rd^2=\f{k^3}{2\pi^2}P_\Rd(k)=\f{1}{8\pi^2\epsilon}\f{H^2}{M_P^2}|\alpha_k+\beta_k|^2=2.4\times10^{-9}}$. Therefore $\ds{\left(\f{M_P}{H}\right)^2\sim5\times10^{6}|\alpha_k+\beta_k|^2/\epsilon}$.}  Using this correspondence the constraint (\ref{eq:ThreeDecades}) becomes,
\begin{equation} \label{eq:MasterConstraint}
10^{3}|\beta|^2\epsilon\ll |\alpha+\beta|^2.
\end{equation}

There are two cases to consider, that of large $\ds{|\beta|}$ and that of small $|\beta|$.  For the case of large $\ds{|\beta|\geq1}$ the inequality restricts $\ds{\epsilon}$ to be very small, $\ds{\epsilon\ll 10^{-3}}$.  The leading contribution to squeezed limit $\ds{f_{NL}}$ in this case is proportional to $\ds{\epsilon(k_1/k_3)}$\cite{Ganc11}, with $\ds{(k_1/k_3)\leq10^3}$.  This results in $\ds{f_{NL}}$, up to numerical coefficients, being forced to be much smaller than order unity, $\ds{f_{NL}\ll\Od\left(1\right)}$.\\

For the case of small $\ds{|\beta|<1}$ the constraint simplifies to $\ds{|\beta|^2\epsilon\ll10^{-3}}$, which upon substitution of (\ref{eq:epsilon}) results in $|\beta|^2\ll10^{-1}$.  As in the mixed case, the largest value of $\epsilon$ is chosen since we would like to maximize $f_{NL}$ which has a leading contribution of the form $\ds{|\beta|\epsilon(k_1/k_3)}$ \cite{Ganc11}.\footnote{Up to numerical coefficients, $\ds{|\beta|^2\epsilon\ll10^{-3}}$ may be written as $\ds{f_{NL}\leq|\beta|\epsilon10^3\ll\sqrt{\epsilon}10^{3/2}}$, which motivates the choice of choosing the largest $\epsilon$.}  This is clearly undetectable as well since we again find, up to numerical coefficients, $\ds{f_{NL}\ll\Od\left(1\right)}$.\\

Therefore inflationary models with all observable modes as excited states do not have a squeezed limit $\ds{f_{NL}}$ which is observationally distinguishable from the Bunch Davies initial state.  We will now parameterize the inequalities used in the proof in order to study its robustness.
\subsection{Unmixed Case: General Parameterization}
In order to develop an intuition for the robustness of the proof previously given, we now repeat the argument directly parameterizing the inequalities.  In general one may take,\\
(1) Subhorizon constraint: $\ds{A_{\text{sub}} H\leq p}$.\\
(2) Backreaction constraint: $\ds{\langle\rho_\Rd\rangle\leq A_{\text{back}}H^2M_P^2}$.\\
(3) $n_{\text{dec}}$ nearly scale invariant observable decades.\\
This parameterization leads to the constraint which is analogous to (\ref{eq:MasterConstraint}) of
\begin{equation}\label{eq:GeneralMasterConstraint}
10^{4n_{\text{dec}}-9}|\beta|^2\epsilon\leq\f{A_{\text{back}}}{A_{\text{sub}}^4}|\alpha+\beta|^2.
\end{equation}
For the case of large $|\beta_k|\geq1$ the constraint (\ref{eq:GeneralMasterConstraint}) simplifies to $\ds{\epsilon\leq10^{9-4n_{\text{dec}}}\left(A_{\text{back}}/A_{\text{sub}}^4\right)}$.  Therefore the leading contribution to $f_{NL}$ is given, up to numerical coefficients, by
\begin{equation}\label{eq:GENERALfNLlarge}
f_{NL}\sim\epsilon(k_1/k_3)\leq10^{12-4n_{\text{dec}}}\left(A_{\text{back}}/A_{\text{sub}}^4\right).\hspace{20mm}(|\beta|\geq1)
\end{equation}

For the case of small $|\beta_k|<1$ the constraint (\ref{eq:GeneralMasterConstraint}) simplifies to $\ds{|\beta|^2\leq10^{11-4n_{\text{dec}}}\left(A_{\text{back}}/A_{\text{sub}}^4\right)}$, where we have once again used ($\ref{eq:epsilon}$) to constrain $\ds{\epsilon\leq10^{-2}}$ and chosen the largest value of $\epsilon$ in order to maximize $f_{NL}$.  Up to numerical coefficients, $f_{NL}$ is then given by
\begin{equation}\label{eq:GENERALfNLsmall}
f_{NL}\sim|\beta|\epsilon(k_1/k_3)\leq\sqrt{10^{13-4n_{\text{dec}}}\left(A_{\text{back}}/A_{\text{sub}}^4\right)}.\hspace{20mm}(|\beta|<1)
\end{equation}

In order to gain some intuition for these numbers, consider the values given in Table 1.  The chosen values for the parameterization variables are very conservative, physically corresponding to modes being barely inside the horizon at the beginning of inflation and the energy density of fluctuations being as large as one tenth the total energy density.  For the number of nearly scale invariant observable decades, we note that three decades is a conservative lower bound \cite{Peiris10} and therefore demonstrate the dependence as we allow more scales to be considered.  The values found for $f_{NL}$ are always less than order unity in both the case of large $|\beta|$ (\ref{eq:GENERALfNLlarge}) and small $|\beta|$ (\ref{eq:GENERALfNLsmall}).
\begin{center}
$\ds{\begin{array}{|lll|c|c|}
\hline
A_{\text{sub}}&A_{\text{back}}&n_\text{dec}&\text{Large }|\beta|\text{ }f_{NL}\text{ }(\ref{eq:GENERALfNLlarge})&\text{Small }|\beta|\text{ }f_{NL}\text{ }(\ref{eq:GENERALfNLsmall})\\\hline
2&0.1&3&\leq\Od\left(10^{-3}\right)&\leq\Od\left(10^{-1}\right)\\
2&0.1&3.5&\leq\Od\left(10^{-5}\right)&\leq\Od\left(10^{-2}\right)\\
2&0.1&4&\leq\Od\left(10^{-7}\right)&\leq\Od\left(10^{-3}\right)\\\hline
\end{array}}$\\
TABLE 1: Typical $f_{NL}$ values for conservative parameterization variables.
\end{center}
Clearly, for any realistic model of inflation one will not be able to generate an observable squeezed limit $f_{NL}$ for models in which all observable modes today are in the excited state.  It should be emphasized that our conclusion of an undetectable squeezed limit $\ds{f_{NL}}$ for the unmixed case required only the four constraints discussed at the beginning of section 2.  Therefore the analysis is very model independent.  However, as a concrete example we will present in the next section a specific model with physically motivated large Bogoliubov coefficients.
\section{Example: Case of Slow Roll Enhanced $\ds{|\beta_k|}$}
\subsection{Review of the Scenario}
We now explicitly show an example to illustrate that inflationary models with excited state vacua that have slow roll enhanced Bogoliubov coefficients cannot have observable consequences today.  These large Bogoliubov coefficients are motivated by allowing for a kinetic dominated era before inflation begins.  Following the model in \cite{Carney12}, expansion of the universe is driven by a single scalar field (the inflaton) minimally coupled to gravity.  The inflationary potential is of the form,
\begin{eqnarray}
    V(\phi)=\left\{
    \begin{array}{llll}
        0 &,& \phi < \phi_i &\text{ (Kinetic Era)}\\
        \Lambda ^4(1 - \alpha \phi) &,& \phi_i < \phi < \phi_f&\text{ (Inflationary Era)}
    \end{array}\right. ,
\end{eqnarray}
where $\ds{\alpha \phi \ll 1}$ for $\ds{\phi_i < \phi < \phi_f}$.  We will assume that post-inflationary physics does not affect our results for the entirety of this study.\\

We are interested in the behavior of the comoving curvature perturbation at the transition point $\ds{\phi_i}$.  In this study we are focused on modes which will potentially be observable today and therefore restrict our attention to classifying modes which are subhorizon at the onset of inflation.  Depending on how the modes experience the transition they can be divided into four regimes according to the physical momentum scale, $\ds{p}$, and characteristic transition time, $\ds{\tau}$:
\begin{center}
$\begin{array}{|lll|}
\hline
H\ll p\ll\tau^{-1}&:&\text{(Sudden Transition)}\\
H\ll p\sim\tau^{-1}&:&\text{(Interpolating Regime)}\\
H\ll \tau^{-1}\ll p&:&\text{(Adiabatic Transition)}\\
M_P\leq p&:&\text{(No Transition)}\\\hline
\end{array}$
\end{center}

The mode function for the kinetic\footnote{The kinetic era mode function is consistent with the form of the Mukhanov variable used in \cite{Contaldi03} motivated by taking the Bunch Davies vacuum for very short wavelengths.} and inflationary periods for modes that undergo a sudden transition was reported in \cite{Carney12},
\begin{eqnarray} \label{eq:modefn}
 \mathcal{R}_k(t) &=& \sqrt{\frac{\pi}{48H}}\, H_0^{(2)}\left(\frac{k}{2aH}\right), \text{ (Kinetic Era),}\cr\cr
 \mathcal{R}_k(\eta) &=& \frac{1}{\sqrt{2k^3}}\frac{H}{\sqrt{2\epsilon}}\left[\alpha(k)\left(1+ik\eta \right)e^{-ik\eta} + \beta(k) \left(1-ik\eta \right)e^{ik\eta} \right], \left(\begin{array}{c}\text{Inflationary Era}\end{array}\right),\cr \cr
 \left(\begin{array}{c}\text{Sudden}\\\text{Transition}\end{array}\right)&\rightarrow&\left\{\begin{array}{lll}
 \alpha (k) &=& \sqrt{\frac{\pi}{48w\epsilon}}e^{-iw}\left[ w\epsilon H_0^{(2)} \Big(\frac{w}{2}\Big) - 3 \left(1+iw \right) H_1^{(2)}\Big(\frac{w}{2}\Big) \right], \cr \cr
 \beta(k) &=&   \sqrt{\frac{\pi}{48w\epsilon}}e^{iw}\left[ -w\epsilon H_0^{(2)} \Big(\frac{w}{2}\Big) + 3 \left(1-iw \right) H_1^{(2)}\Big(\frac{w}{2}\Big) \right].\end{array}\right.
 \end{eqnarray}
Here, H is the Hubble parameter during inflation, $\ds{a}$ is the scale factor with $\ds{a_i}$ as its value at the beginning of inflation, $\ds{w=k/a_iH}$, $\ds{\eta=-1/aH}$, and $\ds{\epsilon=\dot{\phi}^2/2H^2M_P^2}$.  The Hankel functions are denoted by $\ds{H_0^{(2)}}$ and $\ds{H_1^{(2)}}$.  In the subhorizon limit, $\ds{w \gg 1}$, these expressions reduce to,
\begin{eqnarray} \label{eq:modefnmod}
\left| \alpha(k)\right|^2 = \left| \beta(k) \right|^2 = \frac{3}{4}\f{1}{\epsilon}, \cr \cr
\textnormal{Arg} \left[ \frac{\alpha}{\beta} \right] = 2w .
\end{eqnarray}
\subsection{Computing $\ds{f_{NL}}$}
For the sake of clarity, the explicit computation of $\ds{f_{NL}}$ in the standard squeezed limit where we have required all three modes satisfy (\ref{eq:modefnmod}) is located in the appendix.  All observable modes in this case must be in the excited state since otherwise the large $\ds{|\beta_k|}$ would violate the observed power spectrum scale invariance.  Since this is a model with large $\ds{|\beta_k|}$, the result of that computation (\ref{eq:stdfNL}) is an $\ds{f_{NL}}$ that is proportional to the slow roll parameter $\ds{\epsilon}$ as discussed in section 2,
\begin{equation}
f_{NL}=\f{5}{24}\epsilon\left(\f{k_1}{k_3}\right).
\end{equation}
In order to satisfy (\ref{eq:MasterConstraint}) we require that $\ds{\epsilon\ll10^{-3}}$ and therefore at most we have $\ds{f_{NL}\ll10^{-1}}$, which is unobservable.  Therefore this model cannot provide a detectable squeezed limit $\ds{f_{NL}}$ and therefore cannot be experimentally discriminated from the standard Bunch Davies scenario.\footnote{The tensor spectrum would not be affected by the transition to leading order since the total energy density is constant across the transition and therefore would not provide any additional evidence in favor of the pre-inflationary kinetic era.}
\section{Conclusions}
We have argued that in order to be consistent with the nearly scale invariant observable power spectrum, one must require that all decades of observable scale today must either be an excited initial state, or a combination of the Bunch Davies and excited initial states with small $\ds{|\beta_k|}$.  The mixed case of allowing for one observable mode to be Bunch Davies and two of the modes to be excited results in a small contribution to $\ds{f_{NL}}$ for $\ds{|\beta_k|\leq0.01}$ while the contribution for $\ds{0.01<|\beta_k|\leq0.1}$, however, may indeed be order unity.  Allowing for the three decades to be in the same initial excited state imposes strong constraints on $\ds{f_{NL}}$ once both the constraints of subhorizon scale and negligible backreaction are taken into account.  We find in particular that the constraints for small $\ds{|\beta|}$ models result in a heavily suppressed $\ds{|\beta|\epsilon}$ and for large $\ds{|\beta|}
$ models a heavily suppressed $\ds{\epsilon}$.  The 
leading contribution to the squeezed limit $\ds{f_{NL}}$ for each respective case is proportional to these terms, resulting in a prediction that is undetectably small.\\

Therefore most models of single field inflation with excited initial states and negligible superhorizon evolution cannot consistently produce an observably large squeezed limit $\ds{f_{NL}}$ for observable modes.  It is only within the small region of parameter space whereby two modes of the squeezed limit are in an excited state with $\ds{0.01<|\beta_k|<0.1}$ and the third is in the Bunch Davies state that one may obtain an $\ds{f_{NL}}$ of order unity.  If the constraints on the scalar spectral index or the tensor to scalar ratio improve with future data, the constraints on $|\beta_k|$ and $\epsilon$ will respectively become more restrictive and may allow us to rule out detecting non-gaussianity from all excited state inflationary models with negligible superhorizon evolution.\\

\textbf{NOTE:} As we were completing this paper, R. Flauger, D. Green and R. A. Porto informed us of their work \cite{Flauger13}, in which they reach similar conclusions.

\section*{Acknowledgments}
We would like to thank Dan Carney, Anindya Dey, Willy Fischler, Jonathan Ganc, Eiichiro Komatsu and Sandipan Kundu for useful discussions.  We would like to thank Raphael Flauger for helpful comments on an earlier version of this paper.  This material is based upon work supported by the National Science Foundation under Grant Number PHY-0969020.
\section{Appendix A: Computing $\ds{f_{NL}}$ for the Case of Slow Roll Enhanced $\ds{|\beta_k|}$}
To compute the bispectrum up to tree level, we follow the approach in \cite{Maldacena03}. The operator $\ds{\hat{\mathcal{R}}(\mathbf{k},t)}$ is defined the same way as in the standard case, but the mode function (\ref{eq:modefn}) is used instead of the standard Bunch-Davies mode function. The following expression is obtained in the limit where all the $\ds{k_i}$ modes satisfy (\ref{eq:modefnmod}),
\begin{equation} \label{eq:ourbispectrum}
\begin{array}{l}
\langle \mathcal{R}_{\mathbf{k}_1}\mathcal{R}_{\mathbf{k}_2}\mathcal{R}_{\mathbf{k}_3} \rangle = (2\pi)^3 \delta \left( \mathbf{k}_1+\mathbf{k}_2+\mathbf{k}_3 \right)\frac{H^{12}}{\dot{\phi}^8} \frac{1}{\prod\limits_i(2k_i^3)} \left[ A_{k_1k_2k_3}+B_{k_1k_2k_3} \right]\\
A_{k_1k_2k_3} = 36 \left( \prod\limits_a \cos^2 \left(w_a \right)\right) \Bigg[ 2 \frac{\ddot{\phi}}{H \dot{\phi}}\sum\limits_b \frac{k_b^3}{\cos^2\left( w_b\right)}+ \frac{1}{2}\frac{\dot{\phi}^2}{H^2}  \left\lbrace \sum\limits_b \frac{k_b^3}{\cos^2\left( w_b\right)} + \sum\limits_{b \neq c} \frac{k_b k_c^2}{\cos^2\left(w_b\right)} \right\rbrace \Bigg]\\
B_{k_1k_2k_3} = \left[ 9 \frac{\dot{\phi}^2}{H^2} \sum\limits_{a>b}k_a^2k_b^2 \right]\left[ \frac{3+\Sigma_c \cos\left(\theta_t-\theta_c\right)+2\Sigma_c\cos \theta_c}{k_1+k_2+k_3} + P_{k_1k_2k_3} \right]\\
P_{k_1k_2k_3} = \Bigg[ \frac{1+2\cos\theta_3- \cos\left(\theta_1+\theta_2 \right)+ \cos\left(\theta_3-\theta_1 \right) + \cos\left(\theta_3-\theta_2 \right)}{k_1+k_2-k_3} + \, \textnormal{cyclic} \Bigg]
\end{array}.
\end{equation}
Here, $\theta_b = 2w_b = 2k_b/(a_iH)$ and $\theta_t = \theta_1+\theta_2+\theta_3$.\\

The relative magnitude of the bispectrum in comparison to the power spectrum can be written in terms of $\ds{f_{NL}}$ \cite{Baumann09} as,
\begin{eqnarray}
\mathcal{B}_{\mathcal{R}}(k_1, k_2, k_3) = \frac{6}{5}f_{NL}^{local}\times \left[P_{\mathcal{R}}(k_1)P_{\mathcal{R}}(k_2) + P_{\mathcal{R}}(k_2)P_{\mathcal{R}}(k_3) + P_{\mathcal{R}}(k_3)P_{\mathcal{R}}(k_1) \right] . \cr
\end{eqnarray}
For our case, the power spectrum for modes satisfying (\ref{eq:modefnmod}) has been derived in \cite{Carney12} to be,
\begin{eqnarray} \label{eq:PowerSpectrum}
P_{\mathcal{R}}(k) = \frac{1}{2k^3}\frac{3H^2}{2M_P^2\epsilon^2} \cos ^2(w) \, .
\end{eqnarray} 
In the squeezed limit we use the relation,
\begin{eqnarray}
\lim_{k_1 \sim k_2 \gg k_3} \mathcal{B}_{\mathcal{R}}\left( k_1, k_2, k_3 \right) = \frac{12}{5}f_{NL} P_{\mathcal{R}}(k_1)P_{\mathcal{R}}(k_3),
\end{eqnarray}
to obtain the squeezed-limit $\ds{f_{NL}}$. This gives us
\begin{eqnarray} \label{eq:stdfNL}
f_{NL} \left(k_1, k_3\right) &=& \frac{5}{24}\epsilon \left(\frac{k_1}{k_3}\right)\left[1 + f \right]\longrightarrow\frac{5}{24}\epsilon \left(\frac{k_1}{k_3}\right).
\end{eqnarray}
Here, the term $\ds{f}$ is a factor proportional to the sum of cosines of various combinations of $\ds{w_\alpha = k_\alpha/(a_iH)}$, which depends sharply on the choice of modes ($\ds{k}$) and can be taken to average out in measurements to zero.
%

\end{document}